\documentclass[journal, a4paper, twoside, 10pt]{IEEEtran}

\usepackage[utf8]{inputenc}
\usepackage[T1]{fontenc}
\usepackage{capt-of}
\usepackage{cite}
\usepackage{url}
\usepackage{pdfpages}
\usepackage{graphicx}
\newcommand{\topic}{Improving the Usability of Privacy Settings in Facebook}

\title{\topic}

\begin{document}
\author{\IEEEauthorblockN{Thomas Paul  ~ Daniel Puscher  ~ Thorsten Strufe}\\
\IEEEauthorblockA{Peer-to-Peer Networks Group and CASED\\
TU Darmstadt, Germany
}
}
\maketitle

\begin{abstract}
The ever increasing popularity of Facebook and other Online Social Networks has left a wealth of personal and private data on the web, aggregated and readily accessible for broad and automatic retrieval.
Protection from both undesired recipients as well as harvesting through crawlers is implemented by simple access control at the provider, configured by manual authorization through the publishing user.
Several studies demonstrate that standard settings directly cause an unnoticed over-sharing and that the users have trouble understanding and configuring adequate settings.
Using the three simple principles of color coding, ease of access, and application of common practices, we developed a new privacy interface that increases the usability significantly.
The results of our user study underlines the extent of the initial problem and documents that our interface enables faster, more precise authorisation and leads to increased intelligibility.
\end{abstract}

\section{Introduction}
\label{Sec:introduction}
Online Social Networks (OSN) are currently changing the way people interact and arguably represent the most intensively used service on the web.
These services are a platform for users to communicate, to share interests and activities with their friends or anybody on the web, and to subscribe to each other's profiles at ease.
OSN thus contain digital representations of social relations that their users maintain.
They cater for a broad range of users of all ages, cultural, and educational background as well as technical expertise, who utilize them to publish Personally Identifiable Information (PII).

Privacy in this work is defined as the disclosure limitations of shared content to explicitly authorized parties or groups of recipients only.
We don't address confidentiality of data towards the service provider, which instead is assumed to be benign.
Considering the detail and personal character of stored data, privacy is a prevalent requirement for OSN.
Necessary condition to achieve privacy in the context of social networks is, that the subject, the PII is related to, has the possibility to restrict or grant access to the concerning information, and gauge it's configuration.

\begin{figure}[htb]
 \includegraphics[scale=0.287]{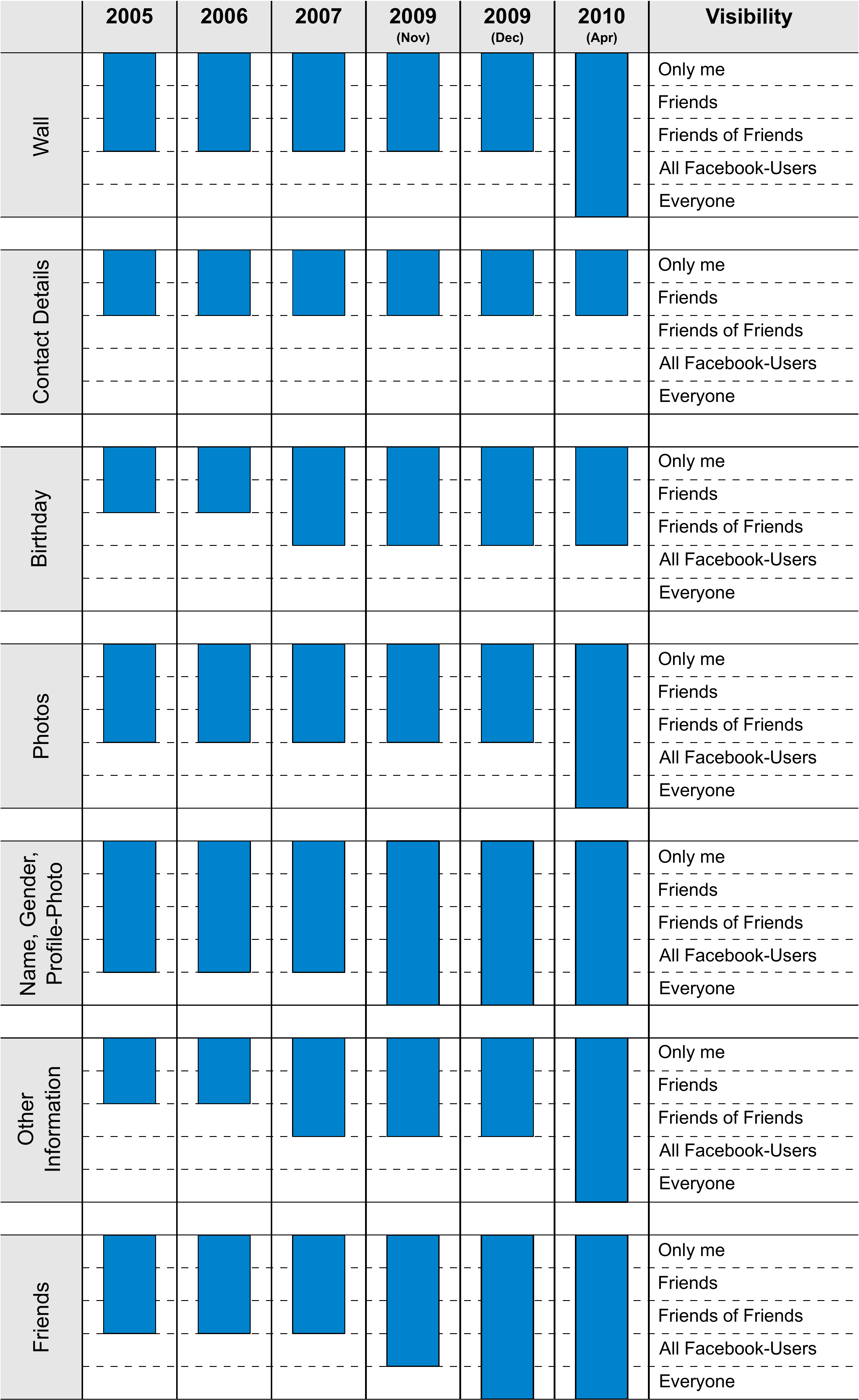}
 \caption{Evolution of Facebook's default privacy settings, cmp. McKeon}
 \label{fig:Development}
\end{figure}

Providers of SNS manage and offer online access to these OSN, which today are based on a client-server approach.
Facebook, catering for over 750 million users, manages the by far biggest Online Social Network.
We hence focus on this service provider and its users.
Facebook provides an interface that enables the users to configure who is allowed to access which information.
The default privacy settings of Facebook entirely expose most parts of the profile and have gradually, yet constantly been relaxed over the last years (cmp. Fig. \ref{fig:Development}).
They allow all members of Facebook and even anybody on the web, regardless if they are registered and authenticated users or not, to collect information about e.g. friends, interests and photos of the users who do not adapt their settings.
This is in stark contrast to strong privacy requirements inherently needed due to the personal character of the shared data.

In order to avoid abuse of personal information, users of the service are required to change the security settings on their own.
Experiments with crawlers \cite{DBLP:journals/corr/abs-1105-6307,bilge09all,strufe10popularity,berjani11recommendation} and several reported incidents show that the average users in fact do not change their security settings.
It is folklore that the main reason for this fact is the low usability of the privacy setting interface.
We hence performed a user study, which demonstrated that the usability of this interface exhibits serious room for improvement.
To this end, in extension to \cite{Paul11improving} we designed a new privacy interface, which is based on the following three criteria:

\emph{Little Effort:}
To ensure a high accuracy when working with the interface, the user should be able check or change his privacy setting with as little effort as possible. 
It is kept as simple as technically feasible, so that it can be understood immediately, even by inexperienced users.

\begin{figure}[htb]
\includegraphics[width=0.45\textwidth]{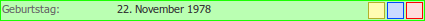}
\caption{Example for an attribute's privacy setting (``Birthday: 22. Nov. 1978)}
\label{fig:Example}
\end{figure}

\emph{Applying common practices} including drag and drop, tooltips or concealing inactive elements help users to easily recognize the current privacy settings and are an instrument to decrease the required effort.
An attribute's visibility for instance can directly be derived from its coloring (cmp. Fig. \ref{fig:Example}).

Results of changes are shown instantly for \emph{direct success control} of each action. 
The used colors are guided by the well-known traffic light colors, adding blue to represent custom settings:
\begin{itemize}
\item \emph{Red:} Visible to nobody
\item \emph{Blue:} Visible to selected friends
\item \emph{Yellow:} Visible to all friends
\item \emph{Green} Visible to everyone
\end{itemize}

We performed a user study comparing the new and the existing interface.
The results demonstrate that the original interface for the privacy settings of Facebook is perceived as confusing by most subjects.
The new interface in comparison was rated as "much better" by all except two and the color coding of the privacy-settings was rated as good to very good by all participating subjects.

The rest of this paper is organized as follows:
we give an overview of the state of the art approaches to privacy protection in online social networks in Section \ref{Sec:sota}, and place it in the context of our requirements and assumptions.
Section \ref{Sec:fb} describes the existing settings and interfaces of Facebook, and in Section \ref{Sec:design} we present the rationale and design of our new interface.
We describe the methodology of our user study in Section \ref{Sec:method} and its results in Section \ref{Sec:evaluation}, before concluding the paper with a summary in Section \ref{Sec:conclusion}.

\section{Related Work}
\label{Sec:sota}
Improving security in OSNs is a very widely discussed issue in literature. 
A  vast amount of approaches have been published, mainly assuming not only malicious users, but a malicious provider, as well. 
The range starts with cutting the profile in centralized OSNs into atomic parts, to encrypt each part separately and distribute keys to authorized recipients \cite{noyb2008,guenther11cryptographic}. It ends with completely distributed p2p OSNs like PeerSoN \cite{buchegger09peerson} or Safebook \cite{cutillo09safebook-2}. 
Common to these approaches is the attempt to enhance service's infrastructure or to even develop an entirely new social network service.
All of them are based on encryption or decentralized storage of private content.
Distrusting the service provider, they consequently aim at implementing distributed access control and confidential data storage.
Assuming benign service providers, the usability of the interfaces emerges as the prevalent challenge for privacy.

The incapability of users to master the settings has been addressed by \cite{Lipford_face}, where a privacy setting interface was presented. which helps users of Facebook to grasp the effect of their changes.
Enforcing grouping of contacts \cite{face_groups} has been proposed as yet another solution to help simplifying the settings by introducing group based authorization.
Direct simplification of the user interface, including color coding for higher intelligibility and 1-click configuration has not previously been proposed, to the best of our knowledge.

\section{Privacy Settings in Facebook}
\label{Sec:fb}
\begin{figure}[H!t]
 \includegraphics[scale=0.46]{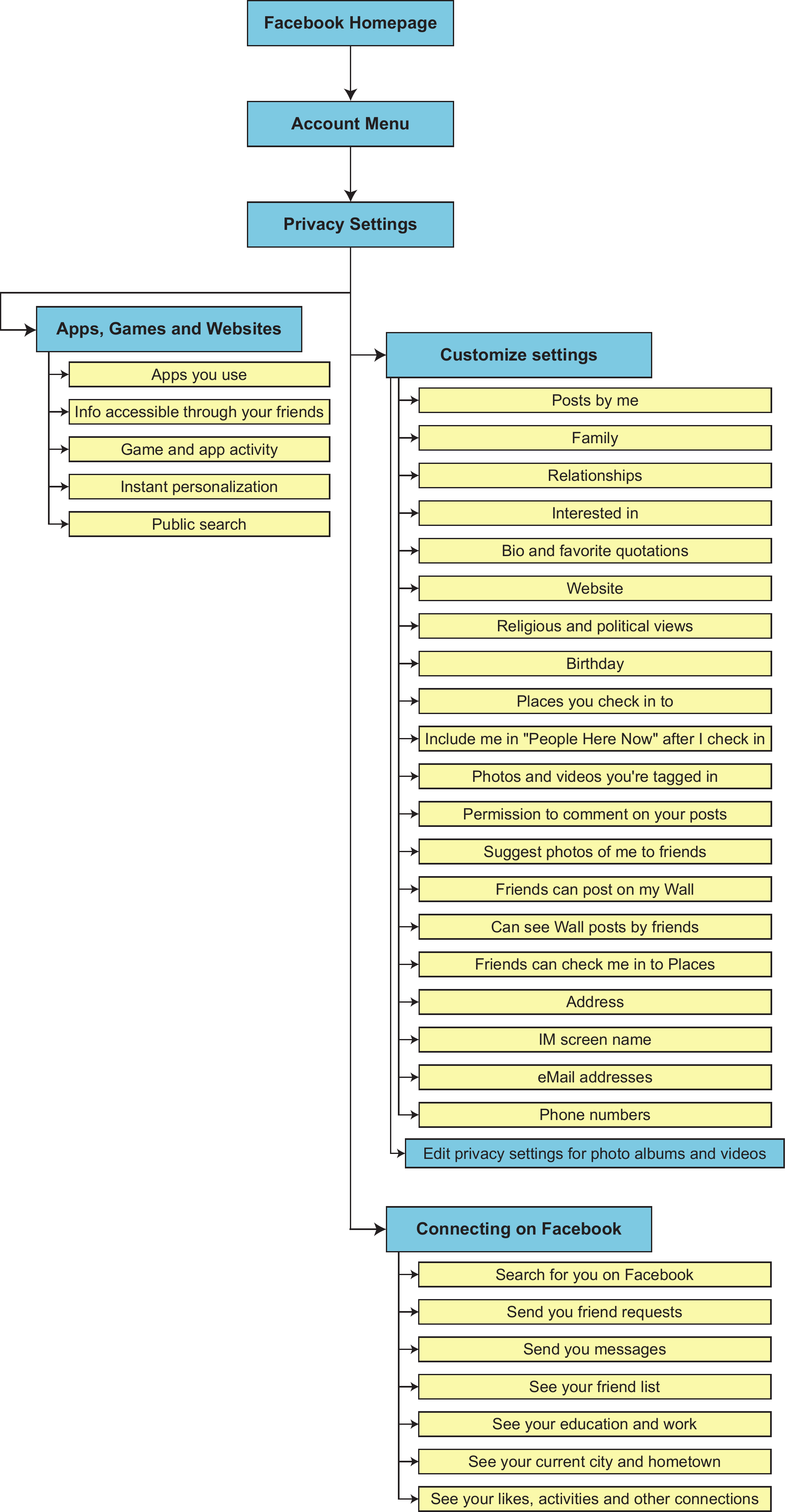}
 \caption{Menu structure of Facebooks privacy settings}
\label{fig:privacysettings}
\end{figure}

Profiles in Facebook consist of several different types of data.
These include identifying information, data on the user's CV, their interests, contacts, and further attributes (cmp. Fig. \ref{fig:privacysettings}).

Facebook offers two possibilities to reach the privacy settings from the entry page. Users have the choice to use the ``My Account'' menu and find the point ``Privacy settings'' or to use the ``profile settings'' in order to reach the privacy settings main page.
It allows to set some preferences on a highly abstract level for categories of profile attributes quickly.
The visibility of those categories can be set to ``Everyone'', ``Friends Only``, ''Friends of Friends`` and ''Recommended``.
The last option represents the default settings (cmp. Fig. \ref{fig:Development}).

These settings, however, effect a small part of the user's profile, which is called ''Sharing on Facebook`` (''Apps, Games and Websites``, Fig. \ref{fig:privacysettings}), only. 
The majority of PII, which is subsumed under  ''Connecting on Facebook``, like photos and albums is not affected by these settings.

On the bottom of the privacy settings main page are links to ''Customize settings``, „Connecting on Facebook“ and to a page where the privacy settings for applications, games and websites may be changed.
''Customize settings`` finally allows for the access authorization to most of the attributes or attribute groups:
\begin{itemize}
  \item Family
  \item Relationships
  \item Interested in
  \item Bio and favorite quotations
  \item Website
  \item Religious and political views
  \item Birthday
  \item Cell phone and other telephone numbers
  \item Address
  \item IM screen name
  \item E-Mail
\end{itemize}

Additional settings, again found at different places on the site, allow for authorizing access to pictures or videos on which the person is explicitly identified ({\em ''tagged``}) by a third party.
The audience that may write messages on a user's wall or to comment on it can be authorized in yet another settings page.
Settings concerning ''Facebook Places``, the location tracking service of Facebook, are located at the same page.
To change the privacy settings of photo albums, it is necessary to click again on another link, which is not highlighted but embedded into a text field.

\section{Improved Interface Design}
\label{Sec:design}
The main design considerations for our new interface where (a) to integrate it into the existing web page of Facebook, to cause the least possible cognitive overhead, and (b) to apply the three criteria mentioned above.

\begin{figure}[htb]
\includegraphics[width=0.45\textwidth]{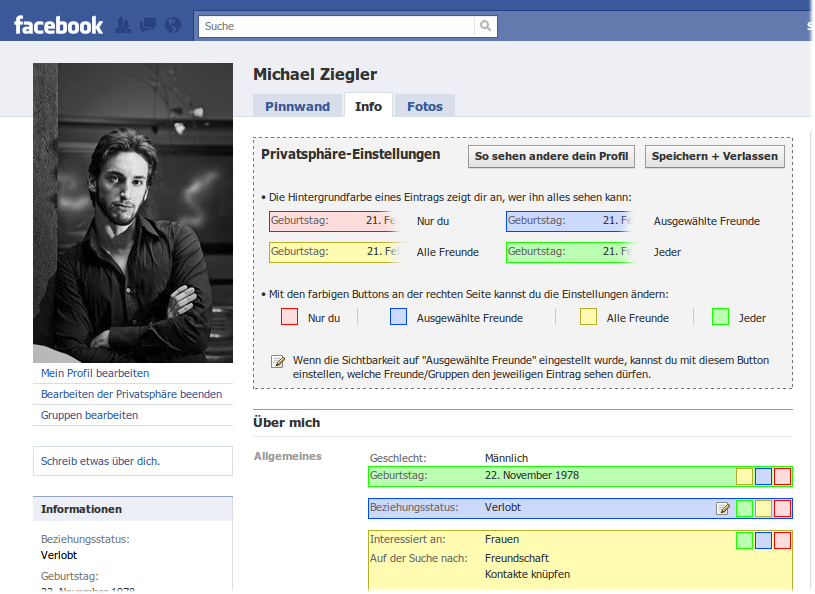}
\caption{Overview of our new interface}
\label{fig:Studyenvironment}
\end{figure}

\subsection{Usage}
\begin{figure}[htb]
\includegraphics[width=0.45\textwidth]{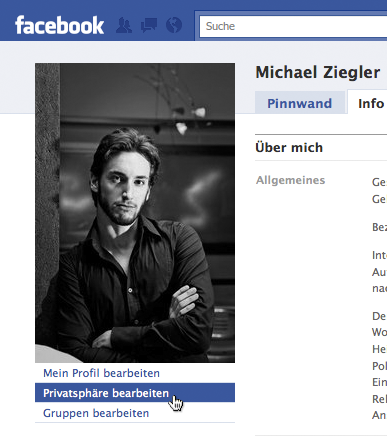}
\caption{Edit Privacy-Settings}
\label{fig:modechange1}
\end{figure}

A prominent link in the main menu, directly underneath the profile picture, switches the profile into configuration mod (''Edit Privacy-Settings``/''Privatsph\"are bearbeiten``, cmp. Fig. \ref{fig:modechange1}).
After editing the privacy settings, the mode can be left by clicking on a link at the same place ("Stop editing privacy settings"/''Bearbeiten der Privatsph\"are beenden``, cmp. Fig. \ref{fig:Studyenvironment}).

All current privacy settings are visualized by a simple coloring scheme in editing mode, as shown above in Fig. \ref{fig:Example}.
The privacy settings of each profile entry can be changed through colored buttons, located directly next to it (cmp. Fig. \ref{fig:Example}).
Tooltips are presented when the mouse hovers over any of the buttons to increase the clarity of the color scheme.

\begin{figure}[htb]
\includegraphics[width=0.45\textwidth]{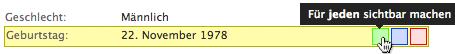}
\caption{Tooltips (''Birthday: 22. November 1978``, with the tooltip ''make visible to {\em everybody}``)}
\label{fig:tooltips}
\end{figure}

Setting the visibility to "anyone", "all friends" or "nobody" happens with just one click on the corresponding button. The settings are changed immediately, which is reflected directly by the change of the color of the cell, containing the considered attribute.

\begin{figure}[htb]
\includegraphics[width=0.45\textwidth]{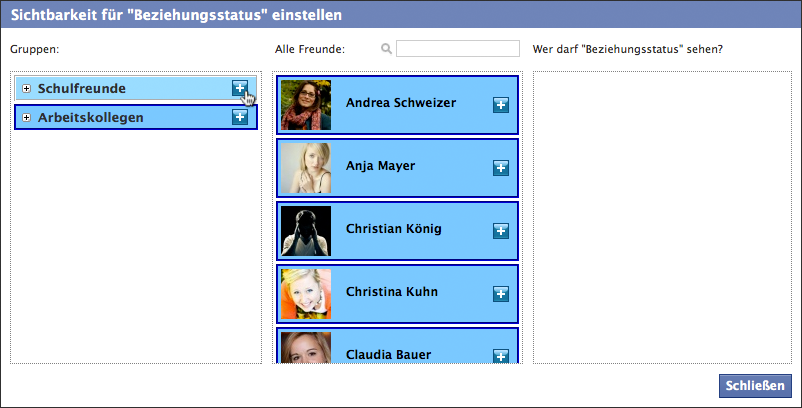}
\caption{User selection with groups}
\label{fig:selectedfriends}
\end{figure}

If the user chooses "selected friends" (blue), a window opens in which friends or groups are granted access to the mentioned attribute.
The window is divided into three columns (Figure \ref{fig:selectedfriends}). The left one contains all groups, the user has created before, the one in the middle shows all friends and the right one is the "visible list" which shows the users that are entitled to see the entry. Adding a friend (from the middle column) or even a whole group of users (from the left column) to this list is done by either clicking on the red ''+`` button or by dragging and dropping the entry into the right column of authorized users. Friends and groups that have been added to the "visible list" are conceiled in the left and the center column.

\begin{figure}[htb]
\includegraphics[width=0.45\textwidth]{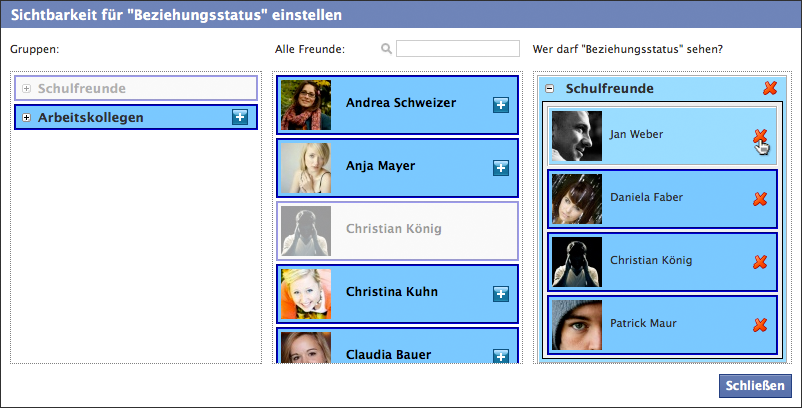}
\caption{Excluding selected users in groups}
\label{fig:selectedfriends2}
\end{figure}

The group in the visible list can be expanded like directories in common file explorers.
Single users or whole groups can hence be granted, or withdrawn access authorization (cmp. Fig. \ref{fig:selectedfriends2}).
Deselecting users only has an effect for the profile entry that the user is editing at that moment. 
The group itself isn't changed.

\begin{figure}[htb]
\includegraphics[width=0.45\textwidth]{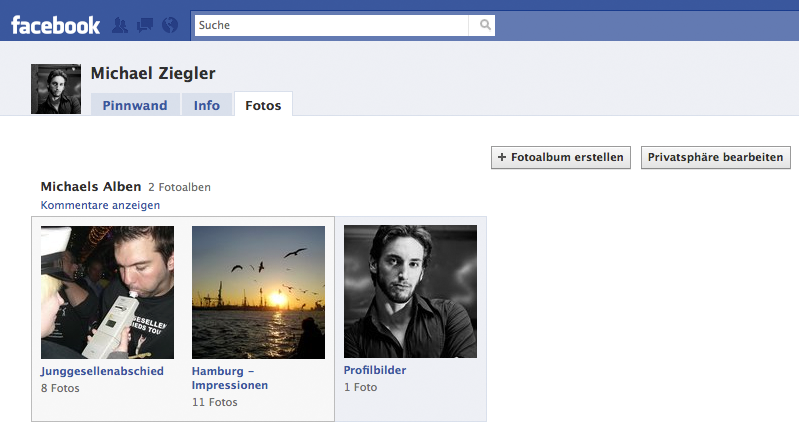}
\caption{Selecting albums}
\label{fig:photos}
\end{figure}

Not only the visibility of the profile, but also the privacy-settings for photo albums can be shown and edited in this way. 
When visiting the "photos" tab in configuration mode, an overview of all photo albums of the user is displayed (cmp. Fig. \ref{fig:photos}).
 
\begin{figure}[htb]
\includegraphics[width=0.45\textwidth]{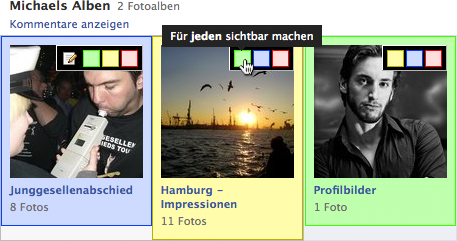}
\caption{Adjusting visibility of photos}
\label{fig:photos2}
\end{figure}

The "Edit Privacy-Settings"-button activates the editing mode according to the profile privacy setting interface and the photo album elements are highlighted with a color. 
The three colored buttons are shown on every item and allow to change the privacy setting like described above (cmp. Fig. \ref{fig:photos2}). 

Creating and managing groups of friends, as mentioned above, allows users to adjust privacy settings more efficiently than selecting each user per profile attribute individually. 
This reflects a community structure of friends in the real world (like "colleagues" or "good friends") and helps to decide quickly which group is allowed to see a certain profile entry. 

\begin{figure}[htb]
\includegraphics[width=0.45\textwidth]{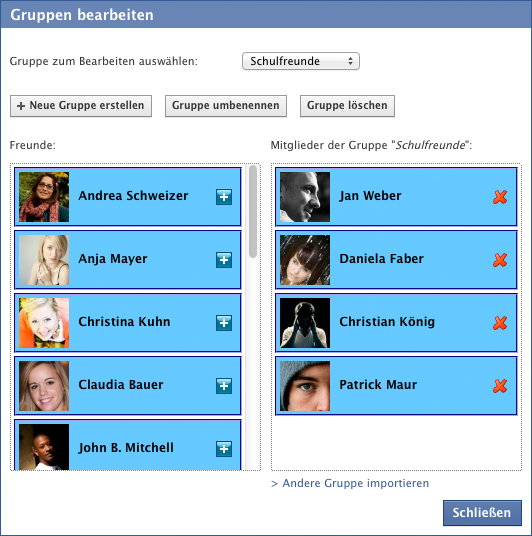}
\caption{Editing Groups}
\label{fig:groups}
\end{figure}

This function can be accessed by an extra button in the left navigation bar and via the account menu. 
The window containing the first part of the group function provides the possibility to select a group, aiming to edit, create, rename or delete a group. 
The next shows two columns. 
The right one displays the members of this group and the left one shows friends that may be added (cmp. Fig. \ref{fig:groups}).  
A friend can be added to a group using drag and drop, or by clicking the "+"-button.
Removal of a friend from a group happens similarly, by clicking the red "X"-button.

\subsection{Prototype Implementation}
We implemented a prototype of our interface as a AJAX/CSS overlay to the original Facebook user profile page, for experimental purposes. 
The goal was seamless integratoin and to change as little as possible when including our enhanced functionality.
PHP and JavaScript were used to realize the visible components, a MySQL database stores necessary data. 
To secure original PII of the test subjects, unnoticed changes to their profiles had to be prevented.
The whole prototype hence was implemented as a mockup, with local hosting of profiles and content.
An Apache HTTP server finally served the whole content for the user study.
The ajax framework Xajax and the JavaScript library script.aculo.us were used for fading elements as well as drag and drop functionality.

The Facebook site was analyzed using the Firefox-Plugin ``Firebug'' and than reengineered for the mockup in order to add our new interface.

Several of the original, partially interleaved CSS style descriptions were merged for higher clarity of the code and some new ones were added for handling the
style of the added components. 
Every privacy-level and the corresponding button has its own style, also the design of the newly added settings windows was realized with CSS-styles. 
All JavaScript specific to Facebook, which was unnecessary for the user study environment, was removed.

\begin{figure*}[t]
 \includegraphics[scale=0.66]{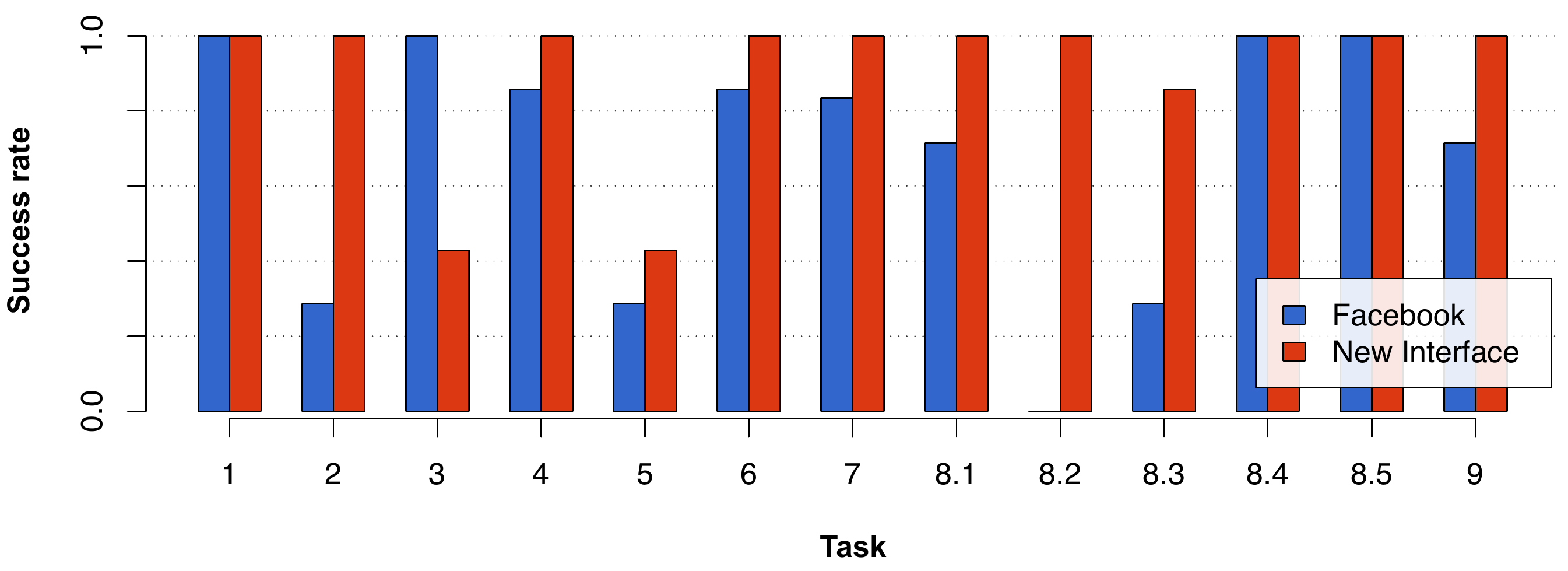}
 \caption{Fraction of successful test persons per task}
 \label{success}
\end{figure*}

To implement the privacy settings, server-side-scripting like storing information into
the MySQL database or reading data from it, was coded with PHP. Xajax
was used to be able to easily execute server generated JavaScript on the client. The interface loads data asynchronously via AJAX. Actions that
are only executed locally in the users browser are written in
Java Script.

\section{Methodology}
\label{Sec:method}
In order to evaluate our solution, four hypothesis have been evaluated in a user study. It was intended to cover
all aspects which may concern users, aiming to adjust their privacy settings.

\begin{itemize}
 \item H1: The new interface makes it easier and faster to find out, to whom a particular attribute is visible.
 \item H2: With the new interface it can be tested quickly, how the complete profile looks like, for another user.
 \item H3: The group management can be handled faster and easier with the new interface.
 \item H4: Setting the visibility of attributes can be realized more effectively with the new interface.
\end{itemize}

\subsection{Sample Description}
The target group for the study consisted of users between 20 and 30 years, because this group represents the majority of users of online social networks. The survey was performed with 20 students aged 20 to 31 years. All test persons are at least member of one online social network. 65\% are visiting these sites at least once a day and 40\% even several times a day. Two thirds of the test persons are Facebook users. The remaining subjects are part of other social networks.

\label{testpersons} Almost all study participants (95\%) have been already in touch with the privacy settings of their network provider. However, many of them call these settings to be confusing (75\%). 20\% of the test persons were very concerned about their privacy settings and stated to modify or check them every month. The rest of them did it less often. 40\% did not change the privacy settings, after they have been set up once.

The possibility to create lists or groups of friends, was used by only 20\% of the participants and the possibility to set certain rights for groups or for individual friends is used by 40\%. When asked if the subjects are aware of  the visibility of their profile's attributes to to other network members, 80\% of them answered "yes".

\subsection {Questionnaire}
The user study was processed by solving some tasks and answering a questionnaire, consisting of four sections: 

\begin{itemize}
 \item General questions regarding the use of online social networks to estimate the prior knowledge of the test person
 \item a practical part where several tasks have to be solved with both interfaces
 \item an evaluation and comparison part of the two interfaces
 \item acquisition of some demographic information like age sex and profession
\end{itemize}

In the practical part of the study, we asked the test persons to

\begin{itemize}
 \item find out to which users or groups the birthday / hometown / relationship status / a photo album is visible
 \item find out which attributes are visible for a specific friend
 \item create a group ``best friends''
 \item add the two friends and the group ``class mates'' to the group ``best friends''
 \item adjust the privacy settings of some attributes and one photo album.
\end{itemize}

To compare the new interface with the interface currently used at Facebook, the tasks of the study was performed with both interfaces. For each task, several measurements were made:

\begin{itemize}
 \item \emph{Time:} Time it takes for a test person to perform a task.
 \item \emph{Hits:} Counted number of clicks it takes a user to complete a task.
\item \emph{Precision:} The  task-solving precision of a study participant. It is only distinguished between the values 1 (task solved completely and correctly) and 0 (task resolved only partially or not at all) because it is very difficult to evaluate the accuracy of the result of a partially solved task with a number between zero and one.
\end{itemize}

\section{Evaluation}
\label{Sec:evaluation}
\begin{figure*}[t]
 \includegraphics[scale=0.5]{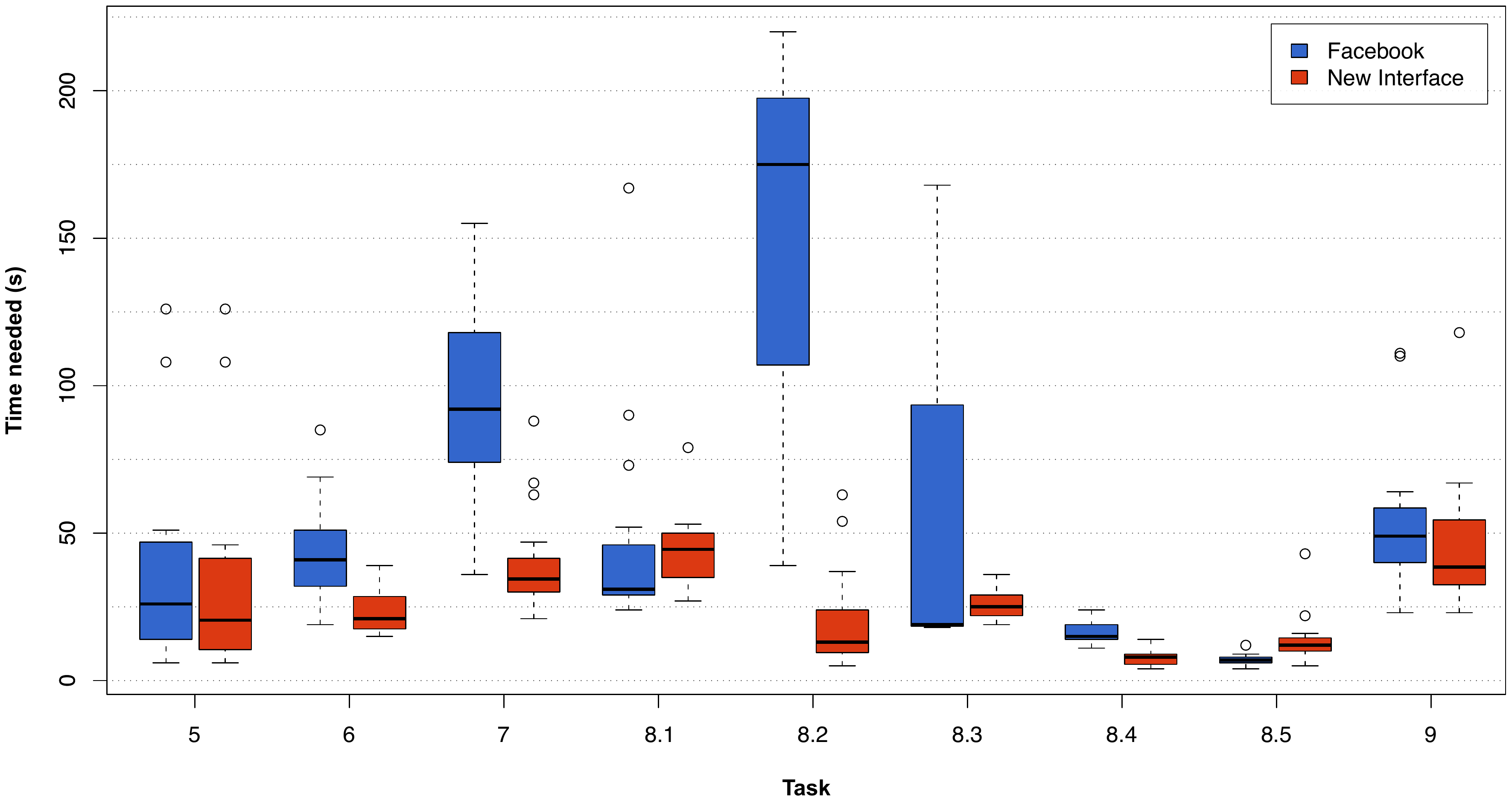}
 \caption{Number of clicks which were needed to solve the tasks}
 \label{timeneeded}
\end{figure*}

Our expected results were, that the tasks can be solved better and faster with the new interface. However, when comparing the precision of the tasks, solved with the new interface and the one, used by Facebook, it is noteworthy that the test persons achieved better results with the new interface (Figure \ref{success}), even if they are Facebook experts, using it every day. The only exception is task 3. The subjects where asked to which users or groups the attribute ``Relationship Status'' is visible. A reason for this exception was not found.

\begin{figure*}[t]
 \includegraphics[scale=0.5]{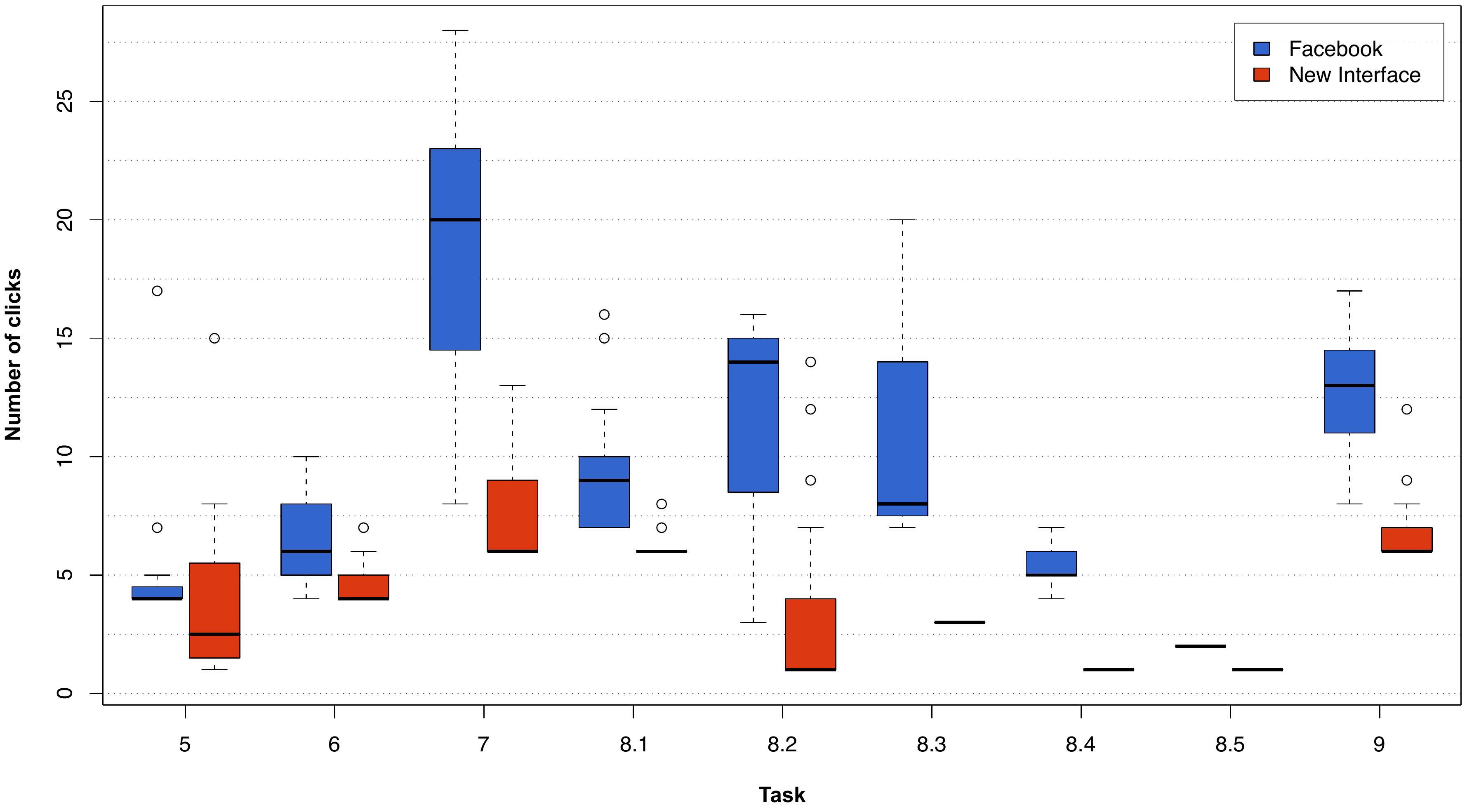}
 \caption{Time it took to solve the tasks}
 \label{numberofclicks}
\end{figure*}

In the first four tasks, subjects had to find out to whom a particular attribute of the profile is visible. The result was that almost all of these tasks could be solved more reliably, using the new interface. The biggest difference could be realized at fields, that are placed on Facebook in the slightly hidden "Connecting on Facebook"-section (task no. 2). The question about the visibility of the field "current city and hometown" on Facebook was answered correctly by only 30\% of participants by using Facebook's privacy page. With the new interface all subjects were able to find out the correct answer.

\subsection{Efficiency Analysis}

\emph{Time needed: }Most tasks can be completed faster by using the new interface (Figure \ref{timeneeded}). Especially when adjusting privacy settings that are in the "Connecting on Facebook"-category (compare Figure \ref{fig:privacysettings}) and while creating groups. On average, the test users need more than twice as much time to solve the tasks with the Facebook interface, compared to the new interface. It is also obvious that for most tasks the new interface has a much smaller variance of results. 

\emph{Clicks needed: } Looking at the number of clicks (Figure \ref{numberofclicks}), the results are very similar to those from the time measurement. When working with the new interface, most tasks can be solved with fewer clicks and the variance is very low.

\subsection{System Usability Scale}
In order to evaluate the usability of the new interface, the standardized questionnaire "System Usability Scale", introduced by Brooke \cite{sus}, was performed. This allows measurements concerning effectiveness, efficiency and user satisfaction. This part was filled out right after using the new interface. It is applicable to various types of systems, because the questions are very general. This allows many different systems, to be compared with each other.

The average SUS-value for our interface (all users), was 87.9. The maximum value was 100, the maximum possible value. The user who rated the interface worst, valued it with 72.5 after all.

Regarding to A. Bangor et al. who analyzed the results of 2324 studies with SUS in the last ten years, acceptable products have a SUS-score of over 70. Better products starting at the high 70s and ending in the upper 80s range. Only truly excellent products have a score above 90. Due to this scale, the usage of our interface is very good.

\begin{figure}[htb]
\includegraphics[width=0.45\textwidth]{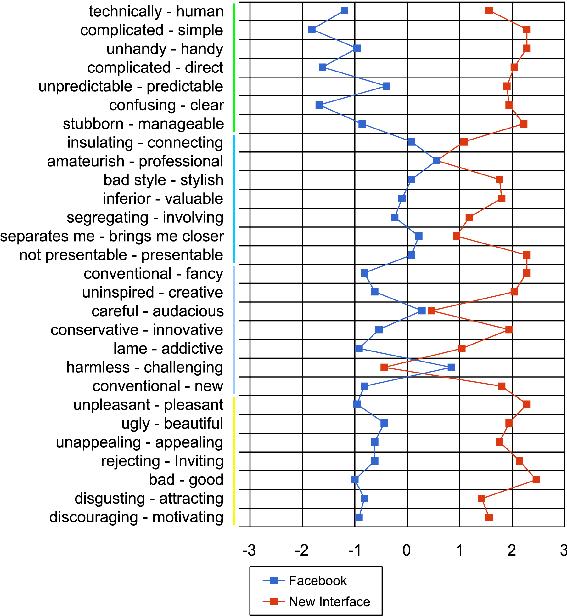}
\caption{Full attrakdiff(tm) evaluation results of the interface}
\label{attrakgdiff}
\end{figure}

\subsection{Comparing users with and without Facebook Accounts }
Since many test persons were users of Facebook, they had advantages while solving the tasks, because they already knew the look and feel of the Facebook site or even the concerning privacy settings. Unsurprisingly, almost all tasks have been solved better by participants that are Facebook users.

Considering only the subjects who do not use Facebook, the difference when handling the two interfaces is bigger. These users coped with the Facebook interface much worse. On average, the test persons solved 90\% of the tasks when using the new interface and only 70\% of them when using Facebook.

\subsection{Identified Problems}
During our study, some floors of our prototype were discovered. We asked the study participants what they would improve. Additionally, we analyzed the videos of the task solving persons in order to find difficult tasks which caused problems in understanding and finding the right solution. In this section we present our findings and ideas for improvement.

\begin{itemize}
 \item \emph{"Selected friends" not completely understandable}

A problem with the new interface occurred for some test persons when dealing with the setting "selected friends" (task no. 3). Thus, a task in which the users first came into contact with this function and should read out the existing settings got solved only by 70\%. When the test persons first actively worked with this function and made settings, there were no further problems.

\smallskip

Some issues in detail:

\begin{itemize}
\item One test person thought that the non-concealed entries in the central column show the friends that are selected.
\item One test person thought that groups in the left column are selected which are not gray.
\item  When adding users, the group that has previously been added was not removed, but the friends should not be added. So the group was still added to the list. While doing an upgrade of the group, other users may get access to the attribute.
\end{itemize}

\emph{Solution:}
To increase the understandability, a help text at the beginning of the window could be inserted, larger column headings or different colors for the different columns would also be a possible solution. For example, the right column could be colored in a way that it generates more attention.

\item \emph{Profile-preview not found:}

In task three we asked, which elements of the profile are visible to a certain friend. This can be done - with reasonable effort - only by using the profile preview. As the caption of the buttons on Facebook is not intuitively understandable ("Preview My Profile"), in the new interface another title has been selected ("How others see your profile"). Nevertheless, only 60\% of the test persons found this function, which is only a minimal increase in comparison to the number of subjects who found it on Facebook (55\%).

\emph{Solution:}
Maybe this function should be pointed out separately. Other options are to reconsider the caption of the button again, or to change the design of the buttons to make them more conspicuous.

\end{itemize}

\section{Conclusion and Future Work}
\label{Sec:conclusion}
This paper deals with the intelligibility and usability of the authorization controls (``privacy settings'') of Facebook.
Analyzing the existing interface, several shortcomings are identified, which we assume lead to the problem of over sharing due to negligent configuration.
This assumption is especially underlined by the fact that an increasing fraction of OSN users are concerned with the privacy settings of the platforms (cmp. Section \ref{testpersons}), leading to the conclusion that users do experience problems in correctly adjusting these settings.

A new interface was introduced, aiming to help ameliorate this situation. 
Based on the three main concepts of color coding, grouping contacts, and proximity of data and controls, it simplifies the interaction and makes the task of authorization as intuitive as possible.
The interface then was implemented as an overlay to the design of Facebook at the time of the study (early 2011).

An extensive user study subsequently was conducted to evaluate its usability.
The subjects were given several tasks, including both authorization as well as revision of current settings, with both the new and the original interface of facebook, consecutively in random order.
The comparison supported that the new interface is easier to use and makes it easier to understand current settings, than the original interface.
The participants of the study were able to solve the tasks much faster and achieved a higher precision, when using our new design.
Facebook has changed the complete user interface in the meantime, and our study hence is not directly applicable to the current state.
The updated design, however, only has a negligible effect on the privacy settings, and we postulate it safe to claim that our results still very well reflect the current situation, none the less.

We currently are in the process of implementing the new user interface as an overlay to the original Facebook site in a Firefox - plugin.
It will allow for easier authorization and lead to higher intelligibility of the privacy controls of Facebook, seamlessly.
We are analyzing the applicability of our main concepts to google+\footnote{\url{http://plus.google.com}}, which already provides much simpler privacy controls as compared to Facebook, at the same time, to identify possible enhancements to both the existing interface as well as our usability concepts.

\section*{ACKNOWLEDGEMENT}
We wish to express our gratitude to Ralf Werner for in-depth discussions, leading us to our design. 
This work in parts has been supported by the IT R$\&$D program of MKE/KEIT of South Korea (10035587, Development of Social TV Service Enabler based on Next Generation IPTV Infrastructure), as well as LOEWE/CASED.

\bibliographystyle{plain}

\newpage

\section{Appendix}
For the purpose of comparison and convenience, we provide the questionnaire together with a translation of the questions as an appendix.

\subsection{Translation of the Tasks}
\begin{enumerate}
  \item Find out, to whom the attribute ``birthday'' is visible.
  \item Find out, to whom the attribute ``current city and hometown'' is 
visible.
  \item Find out, to whom the attribute ``relationships'' is visible.
  \item Find out, to whom the photo album ``stag party'' is visible.
  \item Find out which attributes of your profile are visible for your 
friend ``Stephanie Schmidt''.
  \item Create a list of friends called ``good friends''.
  \item Add all members of the list ``schoolmates'' to the list you just 
created (``good friends''). Then additionally add your Friends ``Claudia 
Bauer'' and ``Sun Yen''.
  \item Change your privacy settings as shown below:
\begin{enumerate}
   \item Cellphone number: Only "Jan Weber" and "Daniela Faber"
   \item Likes, activities and other connections: Everyone
   \item Current city and hometown: Only the list ``schoolmates''
   \item Relationships: Only me
   \item Religious and political views: Friends only
\end{enumerate}
  \item Change the privacy settings of the photo album ``stag party'' so 
that it is visible to the list ``good friends'' but not for ``Patrick 
Maur''.
\end{enumerate}



\section*{Supplementary material}

Proband Lfd. Nr.: _____          Datum: _____

**1.) Für welche Benutzer/Gruppen ist das Attribut „Geburtstag" sichtbar?**

[                                                                    ]

**2.) Für welche Benutzer/Gruppen ist das Attribut „Derzeitiger Wohnort/Heimatstadt" sichtbar?**

[                                                                    ]

**3.) Für welche Benutzer/Gruppen ist das Attribut „Beziehungsstatus" sichtbar?**

[                                                                    ]

**4.) Für welche Benutzer/Gruppen ist das Fotoalbum „Junggesellenabschied" sichtbar?**

[                                                                    ]

**5.) Welche Attribute des Profils sind für den Benutzer „Stephanie Schmidt" sichtbar?**

[                                                                    ]

**6.) Legen Sie eine neue Gruppe mit dem Namen „gute Freunde" an**

**7.) Fügen Sie in die neu erstelle Gruppe „gute Freunde" alle Mitglieder aus der Gruppe „Schulfreunde" ein und zusätzlich noch die beiden Freunde „Claudia Bauer" und „Sun Yen".**

**8.) Stellen Sie die Privatsphäre ihres Profils so ein, dass die Felder wie folgt sichtbar sind:**

- Handy-Nummer:                                  Nur „Jan Weber" und „Daniela Faber"

- „Gefällt mir" und Interessen:                  Jeder

- Derzeitiger Wohnort / Heimatort:              nur die Gruppe „Schulfreunde"

- Beziehungsstatus:                              Niemand

- Religiöse Ansichten / politische Einstellung:  Alle Freunde

**9.) Stellen Sie die Privatsphäre des Fotoalbums „Junggesellenabschied" so ein, dass es von der Gruppe „gute Freunde", allerdings nicht von „Patrick Maur" gesehen werden kann.**

**10.) Beantworten Sie die folgenden Fragen, in dem Sie das entsprechende Kästchen ankreuzen (X).**
Ganz links bedeutet „stimme überhaupt nicht zu" und ganz rechts bedeutet „ich stimme voll und ganz zu". Die Kästchen dazwischen dienen zur Abstufung.

|   |   | Stimme überhaupt nicht zu | | | | Stimme voll und ganz zu |
|---|---|---|---|---|---|---|
| 1. | Ich würde das System gerne öfter benutzen. | ☐ | ☐ | ☐ | ☐ | ☐ |
| 2. | Ich fand das System unnötig kompliziert. | ☐ | ☐ | ☐ | ☐ | ☐ |
| 3. | Das System war einfach zu benutzen. | ☐ | ☐ | ☐ | ☐ | ☐ |
| 4. | Ich denke, dass ich Unterstützung bräuchte, um das System zu benutzen. | ☐ | ☐ | ☐ | ☐ | ☐ |
| 5. | Die verschiedenen Funktionen waren gut in das System integriert. | ☐ | ☐ | ☐ | ☐ | ☐ |
| 6. | Es waren zu viele Unstimmigkeiten im System vorhanden. | ☐ | ☐ | ☐ | ☐ | ☐ |
| 7. | Ich denke, die meisten Leute würden den Umgang mit dem System schnell erlernen. | ☐ | ☐ | ☐ | ☐ | ☐ |
| 8. | Das System ließ sich sehr umständlich benutzen. | ☐ | ☐ | ☐ | ☐ | ☐ |
| 9. | Ich war sehr sicher im Umgang mit dem System. | ☐ | ☐ | ☐ | ☐ | ☐ |
| 10. | Ich musste eine Menge Dinge lernen, bevor ich mit diesem System loslegen konnte. | ☐ | ☐ | ☐ | ☐ | ☐ |



\end{document}